\documentclass[reprint, amsmath,amssymb,aps]{revtex4-2}

\usepackage{graphicx}
\usepackage{dcolumn}
\usepackage{bm}
\usepackage{amsmath}
\usepackage[usenames,dvipsnames]{xcolor}
\usepackage{mathtools}
\usepackage[colorlinks=true,citecolor=blue,linkcolor=blue]{hyperref}
\usepackage[caption=false]{subfig}
\usepackage{bbm}
\usepackage{braket}
\usepackage{comment}
\usepackage{hyperref}
\usepackage{upgreek}
\usepackage{esint}
\usepackage{soul}
\usepackage[normalem]{ ulem } 
\usepackage{floatrow}
\usepackage{verbatim}

\newcommand{\us}[0]{\text{ }\mu\text{s}}

\newcommand{\im}[0]{\text{Im}}

\begin{document}

\title{Robust two-qubit trapped ions gates using spin-dependent squeezing}

\author{Yotam Shapira$^1$}
\author{Sapir Cohen$^2$}
\author{Nitzan Akerman$^1$}
\author{Ady Stern$^2$}
\author{Roee Ozeri$^1$}
\affiliation{\small{$^1$Department of Physics of Complex Systems\\
$^2$Department of Condensed Matter Physics\\
Weizmann Institute of Science, Rehovot 7610001, Israel}}

\begin{abstract}
Entangling gates are an essential component of quantum computers. However, generating high-fidelity gates, in a scalable manner, remains a major challenge in all quantum information processing platforms. Accordingly, improving the fidelity and robustness of these gates has been a research focus in recent years. In trapped ions quantum computers, entangling gates are performed by driving the normal modes of motion of the ion chain, generating a spin-dependent force. Even though there has been significant progress in increasing the robustness and modularity of these gates, they are still sensitive to noise in the intensity of the driving field. Here we supplement the conventional spin-dependent displacement with spin-dependent squeezing, which enables a gate that is robust to deviations in the amplitude of the driving field. We solve the general Hamiltonian and engineer its spectrum analytically. We also endow our gate with other, more conventional, robustness properties, making it resilient to many practical sources of noise and inaccuracies.
\end{abstract}

\maketitle

Two qubit entanglement gates are a crucial component of quantum computing, as they are an essential part of a universal gate set. Moreover, fault-tolerant quantum computing requires gates with fidelities above the fault-tolerance threshold \cite{aharonov2008fault}. Generating high-fidelity two-qubit gates in a robust and scalable manner remains an open challenge, and a research focus, in all current quantum computing platforms \cite{bruzewicz2019trapped,kjaergaard2020superconducting,alexeev2021quantum}.

Trapped ions based quantum computers are a leading quantum computation platform, due to their high controlability, long coherence times and all-to-all qubit connectivity \cite{postler2021demonstration,pino2021demonstration,egan2021fault}. Entanglement gates are typically generated by driving the ions with electromagnetic fields, that creates phonon mediated qubit-qubit interactions. Such gates have been demonstrated, with outstanding fidelities \cite{ballance2016high,gaebler2016high,clark2021high,srinivas2021high}. Moreover, in recent years there have been many theoretical proposals and experimental demonstrations \cite{roos2008ion,green2015phase,haddadfarshi2016high,manovitz2017fast,palmero2017fast,wong2017demonstration,shapira2018robust,zarantonello2018robust,leung2018robust,leung2018entangling,schafer2018fast,webb2018resilient,sutherland2019versatile,figgatt2019parallel,lu2019global,shapira2020theory,sutherland2020laser,lishman2020trapped,wang2020noise,milne2020phase,bentley2020numeric,sameti2021strong,duwe2021numerical,kang2021batch,blumel2021power,blumel2021efficient,dong2021phase,valahu2021robust,wang2022ultra,manovitz2022trapped,fang2022crosstalk,valahu2022quantum} aimed at improving the fidelity, rate, connectability and resilience of such gates. These schemes are largely based on generating spin-dependent displacement forces on the ions which, depending on realization, are linear or quadratic in the driving field. These result in gates which are sensitive to the field amplitude and exhibit a degradation of fidelity which is linear in field intensity noise. A widely used scheme for which is the M\o lmer-S\o rensen (MS) gate \cite{sorensen1999quantum,sorensen2000entanglement}. Driving field amplitude deviations arise naturally in trapped ions systems and may come about due to intensity noise in the drive source, as well as beam pointing noise and polarization noise \cite{brown2016co,mount2016scalable}.

Here we propose a gate scheme which is resilient to deviations in the driving field's amplitude. We combine the conventional spin-dependent displacement with spin-dependent squeezing, by driving the first and second motional sidebands of the ion crystal normal modes. We solve the resulting interaction analytically and formulate constraints on the drive which generate a resilient gate. Crucially, most constraints can be easily satisfied without any numerical optimization. We combine other well-known robustness methods, resulting in a two-qubit entanglement gate which is resilient to many experimental parameters and is independent of the initial motional state, within the Lamb-Dicke regime. Our gates may be implemented using conventional waveform spectral shaping which are straightforward to implement and are common to trapped ions systems. Our method is compatible to laser driven gates as well as laser-free entangling gates \cite{srinivas2021high}.

\begin{figure}
\centering{}\includegraphics[width=1\columnwidth]{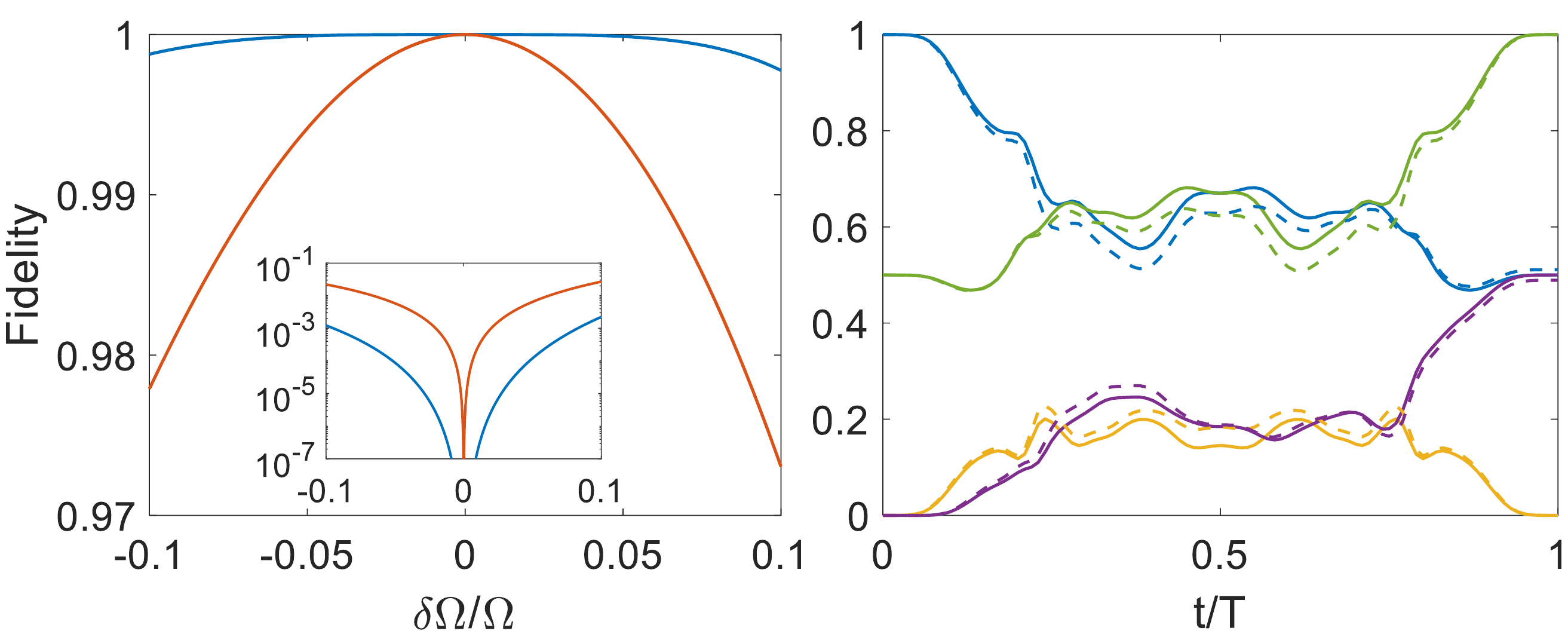} \caption{Robust gate performance. Left: Fidelity of our robust gate (blue) and the conventional MS gate (red), in presence of a deviation of the laser's Rabi frequency, $\delta\Omega$. Our gate shows a flat response, that scales as $\delta\Omega^4$, yielding a high-fidelity operation even in the presence of $10\%$ errors. The MS gate exhibits a quadratic response, and a fast deterioration in fidelity. The inset shows the infidelity, $1-F$, in log scale. Our method typically provides more than two orders of magnitude of improvement throughout the $10\%$ error range. Right: Population dynamics of the initial state $\ket{00}$, for an ideal case (solid) and an erroneous case, with a $5\%$ Rabi frequency error (dashed). In both cases a high fidelity operation (green) is generated at the gate time $t=T$, indicated by an equal population of the $\ket{00}$ (blue) and $\ket{11}$ (purple) states, while the $\ket{01}$ and $\ket{10}$ populations (orange) vanish, indicating robustness to Rabi frequency deviations. The smoothness of the evolution around the gate time is an indication of robustness to gate timing error as well, shown explicitly in the text.}
\label{figMain}
\end{figure}

Figure \ref{figMain} showcases our main results, with the fidelity (left) of our gate (blue) and the conventional MS gate (red), in the presence of deviations in the field's Rabi frequency, $\delta\Omega$. As seen, our gate shows a robust response which scales as $\delta\Omega^4$, and exhibits a high-fidelity entangling operation even with $10\%$ Rabi frqeuency errors. This is contrasted by the quadratic error of an MS gate. The population dynamics of the initial state $\ket{00}$ are shown (right) for the ideal, $\delta\Omega=0$, case (solid) and in presence of a deviation, with $\delta\Omega/\Omega=0.05$ (dashed). While the two scenarios exhibit different dynamics,  at the gate time, $t=T$, they both converge and result in a high-fidelity Bell state (green).

Utilizing spin-dependent squeezing for entangling gates has been suggested in other contexts, such as in order to generate gates in the strong-coupling regime \cite{sameti2021strong}, or in order to generate 3-body \cite{andrade2022engineering} and $n$-body \cite{or2022nbody} interaction terms.

Below we present the Hamiltonian of interest, its solution, the formulation of constraints, and their resolution using spectral shaping. Finally, we analyze our gate's performance and feasibility.

We start with the non-interacting Hamiltonian of two trapped ions, given by,
\begin{equation}
    H_0=\hbar\omega_0 J_z+\hbar\nu a^\dagger a,\label{eqH0}
\end{equation}
with $\omega_0$ the single ion separation frequency of the relevant qubit levels, $J_z=\left(\sigma^z_1+\sigma^z_2\right)/2$ the global Pauli-$z$ operator such that $\sigma^z_n$ is the $z$-Pauli operator acting on the $n$'th ion and $\nu$ the frequency of the center-of-mass normal mode of motion of the ion-chain with its phonon creation operator, $a^\dagger$. All other modes of motion are assumed to be decoupled from the ion's evolution, yet this assumption can be relaxed \cite{shapira2020theory}. The Hamiltonian in Eq. \eqref{eqH0} can trivially be used in a larger ion-chain, by assuming only two ions are illuminated \cite{wang2020high,manovitz2022trapped} and they equally participate in the coupled normal mode.

Without loss of generality and for concreteness we assume the ion qubit levels are coupled by a direct optical transition. The ions are driven by a multi-tone global laser field with a spectral content in the vicinity of the first and second motional sidebands. This yields the interaction Hamiltonian,
\begin{equation}
    V_I=J_x \left[w_1\left(t\right) a^\dagger+ i w_2\left(t\right) \left(a^\dagger\right)^2\right]+H.c,\label{eqVI}
\end{equation}
with $w_n\left(t\right)=\sum_m\rho_{n,m}e^{i\delta_{n,m}t}$. Here  $\rho_{n,m}$ and $\delta_{n,m}$ are amplitudes and frequencies determined below. Equation \eqref{eqVI} is obtained in a frame rotating with respect to $H_0$, and by driving the ion chain with the global time-dependent drive,
\begin{equation}
\begin{split}
    W\left(t\right)=&-\frac{4}{\eta}\sin\left(\omega_0 t\right)\sum_m \rho_{1,m}\cos\left(\left(\nu-\delta_{1,m}\right)t\right) \\ & -\frac{8}{\eta^2} \cos\left(\omega_0 t\right)\sum_m \rho_{2,m}\sin\left(\left(2\nu-\delta_{2,m}\right)t\right),\label{eqW}
\end{split}
\end{equation}
with $\eta$ the Lamb-Dicke parameter, quantifying the coupling between qubit and motional states \cite{wineland1998experimental}. The structure of Eq. \eqref{eqW} implies that the $w_n$'s are proportional to $\Omega$, the driving field's Rabi frequency. The resulting interaction in Eq. \eqref{eqVI} is valid in terms of a rotating wave approximation (RWA) in $\Omega/\omega_0$ and a second order expansion in $\eta$. Furthermore we make use of a RWA in $\Omega/\nu$ allowing us to omit off-resonance carrier coupling terms and counter-rotating terms. Below we incorporate methods that eliminate carrier coupling terms even further \cite{shapira2018robust}. We note that counter-rotating terms still allow for an analytic solution \cite{shapira2020theory}, but are omitted here in favor of a more concise presentation. Note that the $w_n\left(t\right)$'s can be arbitrary complex time-dependent functions.

For the oscillator, the Hamiltonian in Eq. \eqref{eqVI} generates both a spin-dependent displacing term, modulated by $w_1$, and a spin-dependent squeezing term, modulated by $w_2$. In the special case of $w_2=0$ the interaction $V_I$ reduces to the MS Hamiltonian and is exactly solvable. We show below that we may still solve it for non-vanishing second sideband modulations. 

There exists a known solution to general time-dependent quantum harmonic oscillators \cite{harari2011propagator}. However here the appearance of spin-dependence requires special care. We move to a frame rotating with respect to a spin-dependent squeezing by applying a unitary transformation $S\left(J_x r\left(t\right)\right)=\exp\left[\frac{J_x r}{2}\left(a^2-\left(a^\dag\right)^2\right)\right]$, with the time-dependent parameter $r\left(t\right)$, for which we assume $r\left(t=0\right)=0$. This transforms $V_I$ to $V_S=S^\dagger V_I S-i S^\dagger \partial_t S$ (see full expression in section I of the supplemental material). Choosing $w_2\in\mathbb{R}$, i.e. the spectrum of $w_2$ is symmetric around the second sideband, the term in $V_S$ that is proportional to $J_x a^2$ is,
\begin{equation}
    V_{S}^{\left(J_{x}a^{2}\right)}=-i J_{x}a^{2}\left(w_{2}+\frac{1}{2}\partial_{t}r\right).\label{V_Sa2}
\end{equation}

To simplify $V_S$ we are interested in eliminating this term. This yields the trivial differential constraint, $\partial_{t}r=-2w_{2}$, solved by 
\begin{equation}
r=-2\int_0^t dt^\prime w_2\left(t^\prime\right),\label{eqr}
\end{equation}
With these choices, we are left with
\begin{equation}
    V_{S}=J_{x}a\left[w_{1}^{\ast}\cosh\left(J_x r\right)-w_{1}J_{x}\sinh\left(J_x r\right)\right]+H.c\label{eqVS2}
\end{equation}

Since $V_{S}$ in Eq. \eqref{eqVS2} is linear in the mode operators it is analytically solvable. Rotating back to the original frame, the resulting unitary evolution operator due to $V_I$ is,
\begin{equation}
    U_{I}\left(t\right)=S\left(J_x r\left(t\right)\right)D\left(J_x\alpha\left(t\right)\right)e^{-i\left(J_{x}^{2}\left(\Phi_{2}\left(t\right)+\Phi_{4}\left(t\right)\right)+J_{x}\Phi_{3}\left(t\right)\right)}.\label{eqU}
\end{equation}
On the spin side, the evolution in Eq. \eqref{eqU} is composed a global $J_x$ rotation with angle $\Phi_3$, and the desired qubit entangling operation, $J_x^2$ with phase $\Phi_2+\Phi_4$ (Expressions for all $\Phi$'s are given below). On the oscillator side, it is composed of spin-dependent squeezing, $S$ and spin-dependent displacement, $D$, with $D(\alpha)=\exp\left((\alpha a^\dagger-\alpha^\ast a\right)$.

Adopting the useful conventions, $\left\{ f\right\} =\int_{0}^{t}dt_{1}f\left(t_{1}\right)$ and $\left\{ f\left\{ g\right\} \right\} =\int\limits _{0}^{t}dt_{1}\int\limits _{0}^{t_{1}}dt_{2}f\left(t_{1}\right)g\left(t_{2}\right)$, introduced in \cite{sameti2021strong}, $\alpha$ and the $\Phi$'s are given by,
\begin{align}
    \alpha=&\left\{ -i\left(w_{1}\cosh\left(r\right)-J_x w_{1}^{\ast}\sinh\left(r\right)\right)\right\},
    \\
    \Phi_{2}=&\im\left[\left\{ w_{1}^{\ast}\cosh\left(r\right)\left\{ w_{1}\cosh\left(r\right)\right\} \right\} \right],
    \\
    \begin{split}
    \Phi_{3}=&\im\left[\left\{ w_{1}\cosh\left(r\right)\left\{ w_{1}\sinh\left(r\right)\right\} \right\} \right] \\ 
    + & \im\left[\left\{ w_{1}^{\ast}\sinh\left(r\right)\left\{ w_{1}^{\ast}\cosh\left(r\right)\right\} \right\}  \right],
    \end{split}
    \\
    \Phi_{4}=&\im\left[\left\{ w_{1}\sinh\left(r\right)\left\{ w_{1}^{\ast}\sinh\left(r\right)\right\} \right\} \right],\label{eqPhi}
\end{align}

Before analyzing the results in full we note that for a small $w_2$, the leading order contributions to the entangling phase is, $\Phi_2+\Phi_4=\im\left[\left\{ w_{1}^{\ast}\left\{ w_{1}\right\} \right\} +4\left\{ w_{1}\left\{w_{2}\right\}\left\{ w_{1}^{\ast}\left\{w_{2}\right\}\right\} \right\} \right]$, such that $\Phi_2$ scales as $\Omega^2$ and $\Phi_4$ as $\Omega^4$. This dependence is different from that of the MS scheme and its generalizations, and provides the opportunity to mitigate deviations in $\Omega$. 

The form of Eq. \eqref{eqU} allows us to formulate constraints for the generation of two-qubit entangling gates, which are robust to deviations in $\Omega$, and to then choose the proper $w$'s that will satisfy these constraints. We first require that at the gate time $t=T$ there will be no residual displacement or squeezing, i.e., that $r(T)=\alpha(T)=0$, and no rotation of $J_x$, i.e., $\Phi_3(T)=0$. Explicitly, this requires, 
\begin{align}
    \left\{w_{1}\cosh\left(r\right)\right\}=0,\tag{C1}\label{C1}\\
    \left\{w_{1}^{\ast}\sinh\left(r\right)\right\}=0\tag{C2}\label{C2},\\
    r\left(t=T\right)=\left\{w_2\right\}=0,\tag{C3}\label{C3}\\
    \Phi_3\left(T\right)=0\tag{C4}\label{C4}.
\end{align}
Crucially, \eqref{C1} and \eqref{C2} are required to render the gate operation independent of the initial state of the motional mode, i.e. independent of temperature. 

Next, without loss of generality we choose the entanglement phase to be $\varphi=-\pi/2$, a value that rotates the computational basis to fully entangled states,
\begin{equation}\tag{C5}
    \Phi_2\left(T\right)+\Phi_4\left(T\right)=\varphi=-\pi/2.\label{C5}
\end{equation}

Then, robustness to errors in $\Omega$ is provided by,
\begin{equation}\tag{C6}
    \partial_\Omega\left(\Phi_2\left(T\right)+\Phi_4\left(T\right)\right)=0.\label{C6}
\end{equation}
That is, we assume a small error, $\Omega\rightarrow\Omega+\delta\Omega$, and eliminate the leading order contribution of this error to the entanglement phase. This can be generalized to next order terms. In principle similar constraints are required also for other quantities. However, we show below that they are unnecessary by construction.

Our compiled list of six constraints does not uniquely define the drives $w_1,w_2$. We analyze these constraints in terms of frequencies. All the constraints are expressed as integrals from $t=0$ to $t=T$. For these integrals to vanish, the integrands must be composed of non-zero multiples of the gate rate $\xi=2\pi/T$. The choice,

\begin{align}
    w_1\left(t\right)= & \sum_{n} a_{2n+1}e^{i\xi\left(2n+1\right)t},\\
    r\left(t\right)= & \sum_{n} s_{2n}\sin\left(2\xi n t\right),\label{eqSol}
\end{align}
in which $w_1$ is made of odd harmonics of the gate rate and $r$ of a sine series of even harmonics, guarantees that products of the form $w_1\cosh(r)$ and $w_1\sinh(r)$ will not have components at zero frequency, and will therefore integrate to zero. This choice guarantees, then, compatibility with the constraints \eqref{C1}--\eqref{C3}. Furthermore the choice to expand $r$ in a sine series (and not cosine) satisfies \eqref{C4} (see details in section II of the supplemental material). These considerations are independent of $\Omega$ and are therefore resilient to its possible deviations.

We are left with only two constraints, \eqref{C5}, which sets the entangling phase and \eqref{C6}, which makes this phase robust to deviation in $\Omega$. Appropriately, these can be satisfied with only two degrees of freedom. There are infinitely many solutions to these constraints. The simplest uses only $a_3$ and $s_2$ (setting all other $a$'s and $s$'s to zero). This minimal gate scheme is presented in section III of the supplemental material. 

We employ a more elaborate solution, making use of $a_3$, $a_5$, $a_7$, $s_2$ and $s_4$, in order to combine this new result with previously demonstrated robustness properties: mitigation of unwanted off-resonant carrier and sideband couplings, robustness to deviations in the gate time, resilience to phonon mode heating and robustness to motional mode errors  \cite{haddadfarshi2016high,shapira2018robust,webb2018resilient}. These all correspond to constraints which are linear in the $a_n$'s and $s_n$'s and are therefore straightforward to implement (see section IV of the supplemental material for further details). Yielding the drive,
\begin{align}
    w_1\left(t\right)= & \frac{a}{3}\left(3e^{3i\xi t}-10e^{5i\xi t}+7e^{7i\xi t}\right),\\
    r\left(t\right)= & s\left(\sin\left(2\xi t\right)-\frac{1}{2}\sin\left(4\xi t\right)\right).\label{eqSolRobust}
\end{align}

\begin{figure}
\centering{}\includegraphics[width=1\columnwidth]{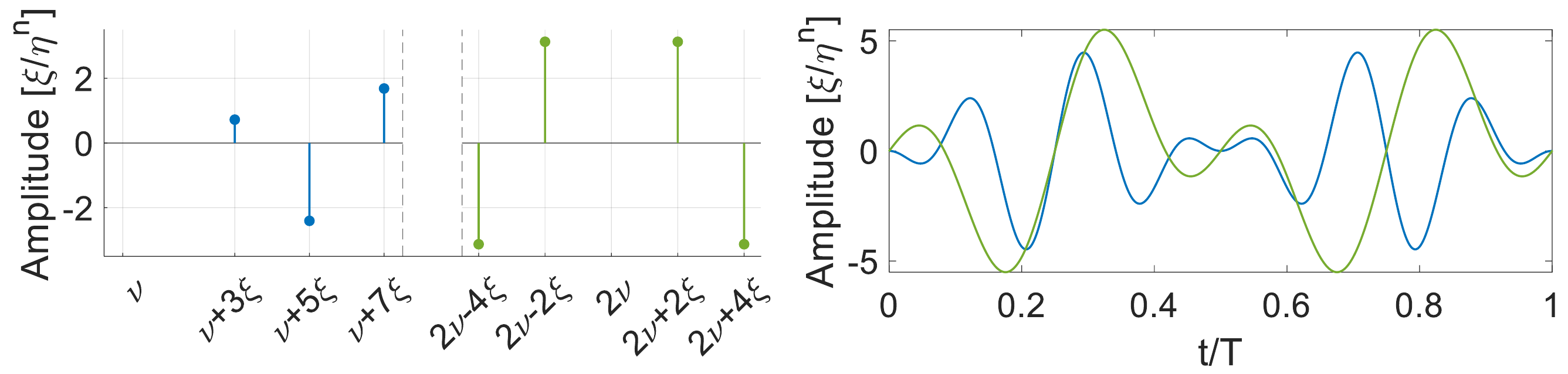} \caption{Spectrum and resulting modulation used to generate our robust gate. Top: Spectrum of the first (blue) and second (green) sidebands. The amplitude is given in units of $\xi/\eta^n$, with $n$ the sideband order. Bottom: Resulting time-domain modulation of the first (blue) and second (green) sidebands. Both modulations vanish continuously at $t=0$ and $t=T$ thus mitigating off-resonance coupling to unwanted transitions.}
\label{figSpect}
\end{figure}

Thus we still have only two undetermined degrees of freedom, $a$ and $s$, in order to satisfy , \eqref{C5} and \eqref{C6}. We progress by Taylor expanding the hyperbolic functions in these constraints to $6$'th order, yielding polynomial equations for $a$ and $s$, which are solved analytically. We then optimize these solutions numerically by directly evaluating \eqref{C5} and \eqref{C6} with a straightforward gradient descent. This yields, $a=0.3608\cdot\xi$ and $s=0.7820$, which constitutes a $3\%$ correction to the analytical solution (see further information in section V of the supplemental material). 

The Rabi frequency required by our scheme is $\Omega_\text{robust}\approx\left(3/\eta+6/\eta^2\right)\xi$. It is more demanding than the MS Rabi frequency, $\Omega_\text{MS}=\frac{\xi}{2\eta}$, showing that robustness is afforded at the price of additional drive power. In the case of two ions, with COM mode $\eta=0.144$, a gate time of $50\us$ requires total laser power of $1.6\text{ mW}$ in the usual MS gate and a power of $150\text{ mW}$ for our fully robust gate (see further details in section VI of the supplemental material). Nevertheless in most implementations the limit on two-qubit entangling gates is fidelity and not driving power. More sophisticated solutions, using additional driving tones can divert field amplitudes from $w_2$ to $w_1$, which reduces the required power but increases the drive complexity. We note that the second sideband drive is significant and cannot be naively treated perturbatively \cite{sameti2021strong}.

The resulting spectrum is presented in Fig. \ref{figSpect} (top) showing the spectral components modulating the first (blue) and second (green) sidebands. The amplitude of the spectrum is normalized by the Lamb-Dicke parameter to the power of the sideband order, i.e. the second sideband modulation is $\eta$ times stronger than the first sideband modulation. The corresponding time-domain modulation of the sidebands due to our drive is shown in Fig. \ref{figSpect} (bottom). We note that both modulations continuously vanish at the start and the end of the gate, which acts to reduce off-resonance coupling to unwanted transitions.

The phase space trajectories generated by our scheme are deduced by the squeezing, $S\left(J_x r\right)$, and displacement, $D\left(J_x \alpha\right)$ operators, in Eq. \eqref{eqU}. We note that due to the appearance of $J_x$ in $D$ and $S$, the states $\ket{++}$ and $\ket{--}$ follow different phase space trajectories. This also occurs in the MS gate, however here we also have $J_x$ operators in $\alpha$, hence the trajectories are not simply reflected about the origin, as in the MS case. We calculate the phase space trajectories by using the identity, $S\left(r\right)D\left(\alpha\right)=D\left(\gamma\right)S\left(r\right)$, with $\gamma=\alpha\cosh\left(r\right)-\alpha^\ast\sinh\left(r\right)$, such that the state is first squeezed by $r$ and then displaced by $\gamma$. We note that the MS phase-space intuition for spin-dependent displacement gate is not valid here, i.e. the entangling phase is not proportional to the area enclosed by the phase space trajectory.

\begin{figure}
\centering{}\includegraphics[width=1\columnwidth]{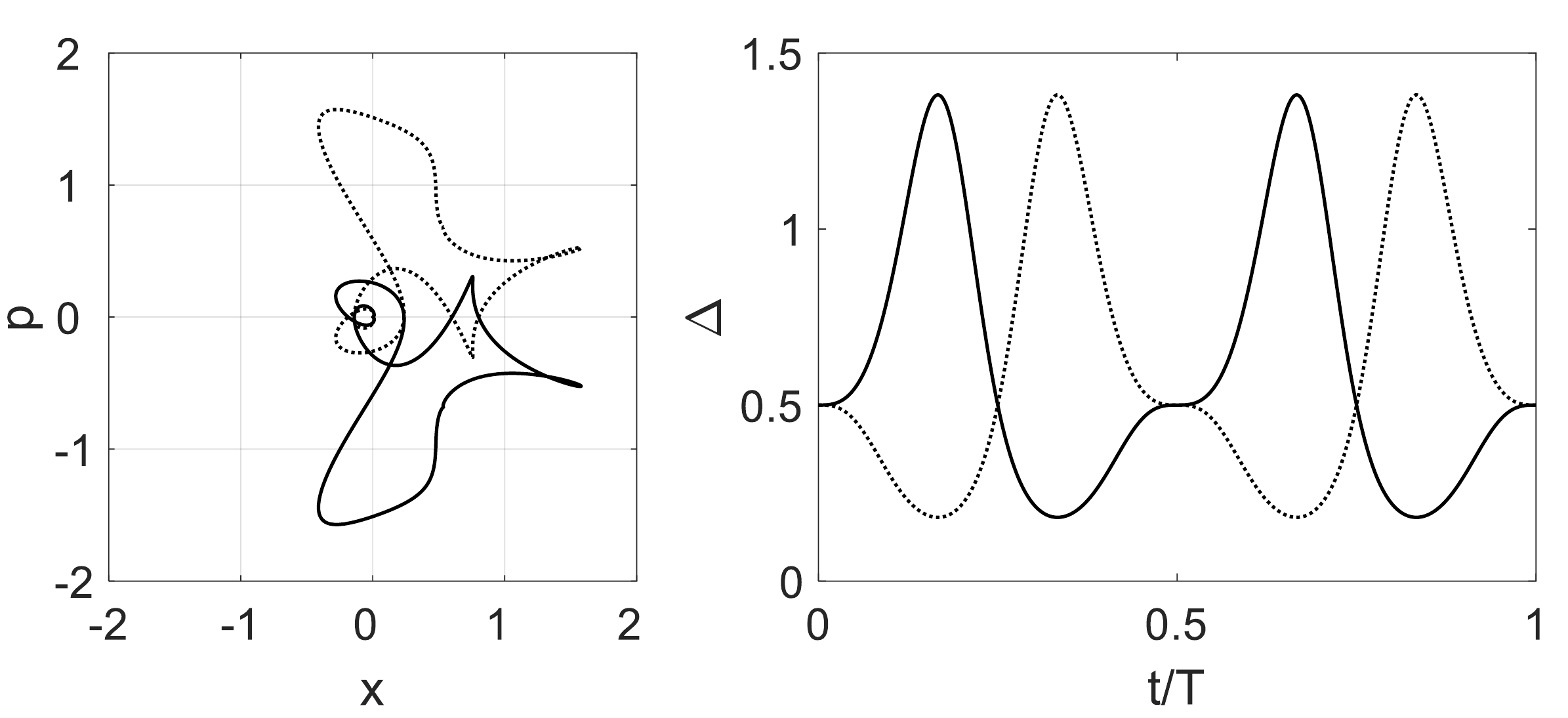} \caption{
Motion in phase space Left: Phase space displacement of the $\ket{++}$ (solid) and $\ket{--}$ (dashed) states. The two trajectories are related by a reflection around the $x$ axis and time reversal. Right: standard deviation of position, $\Delta x=\exp\left(r\right)/2$ (solid) and momentum, $\Delta p=\exp\left(-r\right)/2$ (dashed) for the $\ket{++}$, revealing several oscillations of squeezing and anti-squeezing in both quadratures. These are the same for momentum (solid) and position (dashed) in the $\ket{--}$ state respectively. Since the squeezing parameter $r$ is real then together with displacement (left) the motion in phase space is completely defined.}
\label{figPhase}
\end{figure}

The phase space displacement of $\ket{++}$ is presented in Fig. \ref{figPhase} (left, solid). The same evolution is shown for $\ket{--}$ (dashed). The trajectories are reflected around the $x$ axis and time reversed. Indeed, using $r\left(T-t\right)=r\left(t\right)$ and $w_1\left(T-t\right)=w_1^\ast\left(t\right)$, this is readily confirmed. Squeezing by $r$ changes the expectation value error of position and momentum, $\Delta x$ and $\Delta p$, to $e^r/2$ and $e^{-r}/2$ respectively. Indeed, the figure also shows the standard deviations (right) of $x$ (solid), and $p$ (dashed) for the $\ket{++}$ state, exhibiting non-trivial dynamics. Since $r$ is real the displacement and standard deviations along both axes completely define the phase-space motion.

The form of the evolution operator in Eq. \eqref{eqU}, together with known phase-space identities \cite{gerry2004introductory}, allow us to calculate the gate fidelity. Specifically, we calculate the overlap of the state generated by our gate with the ideal case, assuming the initial state is $\ket{00}$, at the motional ground state (see details in section VII of the supplemental material). This is used to calculate the gate fidelity in presence of Rabi frequency deviations, $\delta\Omega$ shown in Fig. \ref{figMain} (left). 

\begin{figure}
\centering{}\includegraphics[width=1\columnwidth]{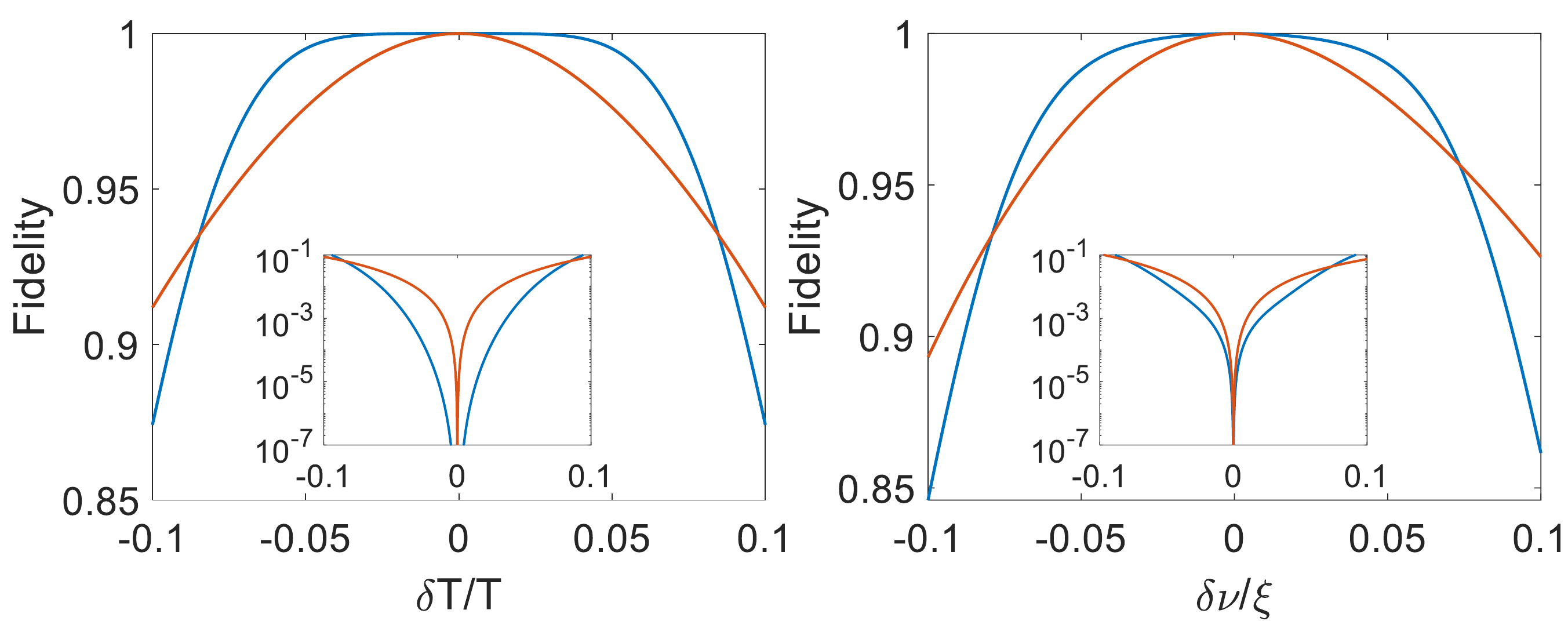} \caption{Additional robustness properties of our gate (blue) compared to the MS gate (red). The insets show the same data recast as infidelity in log scale. Left: Fidelity in presence of gate timing errors, $\delta T/T$. Our gate shows a fourth order, wide response enabling high-fidelity operation even in presence of $5\%$ errors. Right: Fidelity in presence of motional mode frequency errors, $\delta\nu/\xi$. Similarly, our gate exhibits a wide high fidelity region. As expected, the fidelity is not symmetric around the peak.}
\label{figFid2}
\end{figure}

Moreover, as previously demonstrated, the form of the drive in Eq. \eqref{eqSolRobust} ensures that our gate is robust to additional errors and noise. Indeed, Fig. \ref{figFid2} shows our gate fidelity in the presence of gate time deviations, $\delta T$ (left), and motional mode frequency errors, $\delta\nu$ (right). For both of these errors our gate exhibits high fidelity (blue) which scales favorably compared to the MS gate (red).

In conclusion, we have used spin-dependent squeezing in order to propose a two-qubit entangling gate for trapped ions qubits, which is resilient to deviations in the driving field intensity. We have also supplemented our gate with more conventional, previously demonstrated, robustness properties, making it also robust to gate timing errors, motional mode heating, secular frequency drifts, as well as mitigation of coupling to unwanted transitions such as the carrier transition and off resonance sidebands. We do so by generating constraints, which can then be satisfied, almost entirely, with spectral consideration in an analytic fashion. Our new gate can be readily incorporated in the trapped ion quantum toolbox.

We thank David Schwerdt for helpful discussions. This work was supported by the Israeli Science Foundation, the Israeli Ministry of Science Technology and Space, the Minerva Stiftung, the European Union’s Horizon 2020 research and innovation programme (Grant Agreement LEGOTOP No. 788715), the DFG (CRC/Transregio 183, EI 519/7-1), ISF Quantum Science and Technology (2074/19).

\bibliography{main}

\end{document}


\title{Robust two-qubit trapped ions gates using spin-dependent squeezing - Supplemental material}

\author{Yotam Shapira$^1$}
\author{Sapir Cohen$^2$}
\author{Nitzan Akerman$^1$}
\author{Ady Stern$^2$}
\author{Roee Ozeri$^1$}
\affiliation{\small{$^1$Department of Physics of Complex Systems\\
$^2$Department of Condensed Matter Physics\\
Weizmann Institute of Science, Rehovot 7610001, Israel}}

\maketitle
\onecolumngrid
\section{Expression for $V_S$}
We give the full expression of $V_S$ before any assumption is placed on $w_1$ and $w_2$. Using the relation, $V_S=S^\dagger V_I S-i S^\dagger \partial_t S$, we have,
\begin{equation}
\begin{split}
    V_{S} &	=J_{x}a\left[w_{1}^{\ast}\cosh\left(J_{x}r\right)-w_{1}\sinh\left(J_{x}r\right)\right]\\
	&+J_{x}a^{2}\left[-i w_{2}^{\ast}\cosh^{2}\left(J_{x}r\right)+i w_{2}\sinh^{2}\left(J_{x}r\right)-\frac{i}{2}\partial_{t}r\right]\\
	&+J_{x}\left(2a^{\dag}a+1\right)i w_{2}^{\ast}\cosh\left(J_{x}r\right)\sinh\left(J_{x}r\right)+H.c.
\end{split}\label{eqVsFull}
\end{equation}
Our choice (4) of the main text for a real $w_2$ and for its relation to $r$ makes the last two lines of \eqref{eqVsFull} vanish. 

We note that for two qubits we have the useful identities, $J_x^{2n}=J_x^2$ and $J_x^{2n+1}=J_x$, for $n=1,2,...$, such that $J_x^2\cosh\left(J_x r\right)=J_x^2\cosh\left(r\right)$ and $\sinh\left(J_x r\right)=J_x\sinh\left(r\right)$. These are specifically used for deriving Eq. (7)-(11) of the main text.

\section{Satisfying $\Phi_3=0$}
We show that under the general drive, in Eq. (13) of the main text, the constraint (C3) is satisfied, namely, $\Phi_{3}=0$. We have,
\begin{align}
\Phi_{3}&=\im\left[\left\{ w_{1}\cosh\left(r\right)\left\{ w_{1}\sinh\left(r\right)\right\} \right\} +\left\{ w_{1}^{\ast}\sinh\left(r\right)\left\{ w_{1}^{\ast}\cosh\left(r\right)\right\} \right\} \right]	
\end{align}

Focusing on the first term,
\begin{align}
\im\left\{ w_{1}\cosh\left(r\right)\left\{ w_{1}\sinh\left(r\right)\right\} \right\} 	=& \im\int\limits _{0}^{T}dtw_{1}\left(t\right)\cosh\left(r\left(t\right)\right)\int\limits _{0}^{t}dt^{\prime}w_{1}\left(t^{\prime}\right)\sinh\left(r\left(t^{\prime}\right)\right)\\
\begin{split}
=&\im\left(\int\limits _{0}^{T/2}dt\int\limits _{0}^{t}dt^{\prime}+\int\limits _{T/2}^{T}dt\int\limits _{0}^{t-T/2}dt^{\prime}+\int\limits _{T/2}^{T}dt\int\limits _{t-T/2}^{T/2}dt^{\prime}+\int\limits _{T/2}^{T}dt\int\limits _{T/2}^{t}dt^{\prime}\right)\\
&\cdot w_{1}\left(t\right)\sinh\left(r\left(t\right)\right)w_{1}\left(t^{\prime}\right)\cosh\left(r\left(t^{\prime}\right)\right)
\end{split}
\\
\equiv & I_{1}^{\left(1\right)}+I_{2}^{\left(1\right)}+I_{3}^{\left(1\right)}+I_{4}^{\left(1\right)}.
\end{align}
We note that,
\begin{align}
I_{2}^{\left(1\right)}	=&\im\int\limits _{T/2}^{T}dt\int\limits _{0}^{t-T/2}dt^{\prime}w_{1}\left(t\right)\sinh\left(r\left(t\right)\right)w_{1}\left(t^{\prime}\right)\cosh\left(r\left(t^{\prime}\right)\right)\\
=&\im\int\limits _{0}^{T/2}dt\int\limits _{0}^{t}dt^{\prime}w_{1}\left(t+T/2\right)\sinh\left(r\left(t+T/2\right)\right)w_{1}\left(t^{\prime}\right)\cosh\left(r\left(t^{\prime}\right)\right)\\
=&-\im\int\limits _{0}^{T/2}dt\int\limits _{0}^{t}dt^{\prime}w_{1}\left(t\right)\sinh\left(r\left(t\right)\right)w_{1}\left(t^{\prime}\right)\cosh\left(r\left(t^{\prime}\right)\right)\\
=&-I_{1}^{\left(1\right)},
\end{align}
where we used,
\begin{align}
r\left(t+\frac{T}{2}\right)&=\sum_{n}s_{2n}\sin\left(2n\xi\left(t+\frac{T}{2}\right)\right)=\sum_{n}s_{2n}\left(2n\xi t\right)=r\left(t\right),\\
w_{1}\left(t+\frac{T}{2}\right)&=\sum_{n}a_{2n+1}e^{\left(2n+1\right)i\xi\left(t+\frac{T}{2}\right)}=-\sum_{n}a_{2n+1}e^{\left(2n+1\right)i\xi t}=-w_{1}.
\end{align}

Similarly,
\begin{align}
I_{3}=&\im\int\limits _{T/2}^{T}dt\int\limits _{t-T/2}^{T/2}dt^{\prime}w_{1}\left(t\right)\sinh\left(r\left(t\right)\right)w_{1}\left(t^{\prime}\right)\cosh\left(r\left(t^{\prime}\right)\right)\\
=&\im\int\limits _{T/2}^{T}dt\int\limits _{T-t}^{T/2}dt^{\prime}w_{1}\left(\frac{3}{2}T-t\right)\sinh\left(r\left(\frac{3}{2}T-t\right)\right)w_{1}\left(t^{\prime}\right)\cosh\left(r\left(t^{\prime}\right)\right)\\
=&\im\int\limits _{T/2}^{T}dt\int\limits _{T/2}^{t}dt^{\prime}w_{1}\left(\frac{3}{2}T-t\right)\sinh\left(r\left(\frac{3}{2}T-t\right)\right)w_{1}\left(T-t^{\prime}\right)\cosh\left(r\left(T-t^{\prime}\right)\right)\\
=&\im\int\limits _{T/2}^{T}dt\int\limits _{T/2}^{t}dt^{\prime}w_{1}^{\ast}\left(t\right)\sinh\left(r\left(t\right)\right)w_{1}^{\ast}\left(t\right)\cosh\left(r\left(t\right)\right)\\
=&-\im\int\limits _{T/2}^{T}dt\int\limits _{T/2}^{t}dt^{\prime}w_{1}\left(t\right)\sinh\left(r\left(t\right)\right)w_{1}\left(t\right)\cosh\left(r\left(t\right)\right) \\
=&-I_{4}.
\end{align}
where we used the parameter transformations $t\rightarrow\frac{3}{2}T-t$, and $t^{\prime}\rightarrow T-t^{\prime}$, and,
\begin{align}
r\left(\frac{3}{2}T-t\right)	=&r\left(T-t\right)=-r\left(t\right)\\
w_{1}\left(\frac{3}{2}T-t\right)	=&-w_{1}^{\ast}\left(t\right)
w_{1}\left(T-t\right)	=w_{1}^{\ast}\left(t\right).
\end{align}
Thus the first term in $\Phi_{3}$ vanishes. The second term vanishes similarly. We note that these consideration would not have worked out for a cosine spectral decomposition of r.

\section{Minimal $\Omega$-robust gate}

We first generate a minimial gate, abiding constraints (C1)-(C6) without attempting to guarantee any additional robustness properties. This can be obtained by choosing only two degrees of freedom. 

By repeating the process described in the main text we obtain, $a_{3}\approx-1.1521$ and $s_{2}=-0.8896$ such that the required Rabi frequency is $\Omega_{\text{minimal}}\approx\left(\frac{2.3}{\eta}+\frac{1.8}{\eta^{2}}\right)\xi$. This constitutes only a small power advantage compared to the fully robust gate described in the main text and in the next subsection.

\begin{figure}
\centering{}\includegraphics[width=0.5\columnwidth]{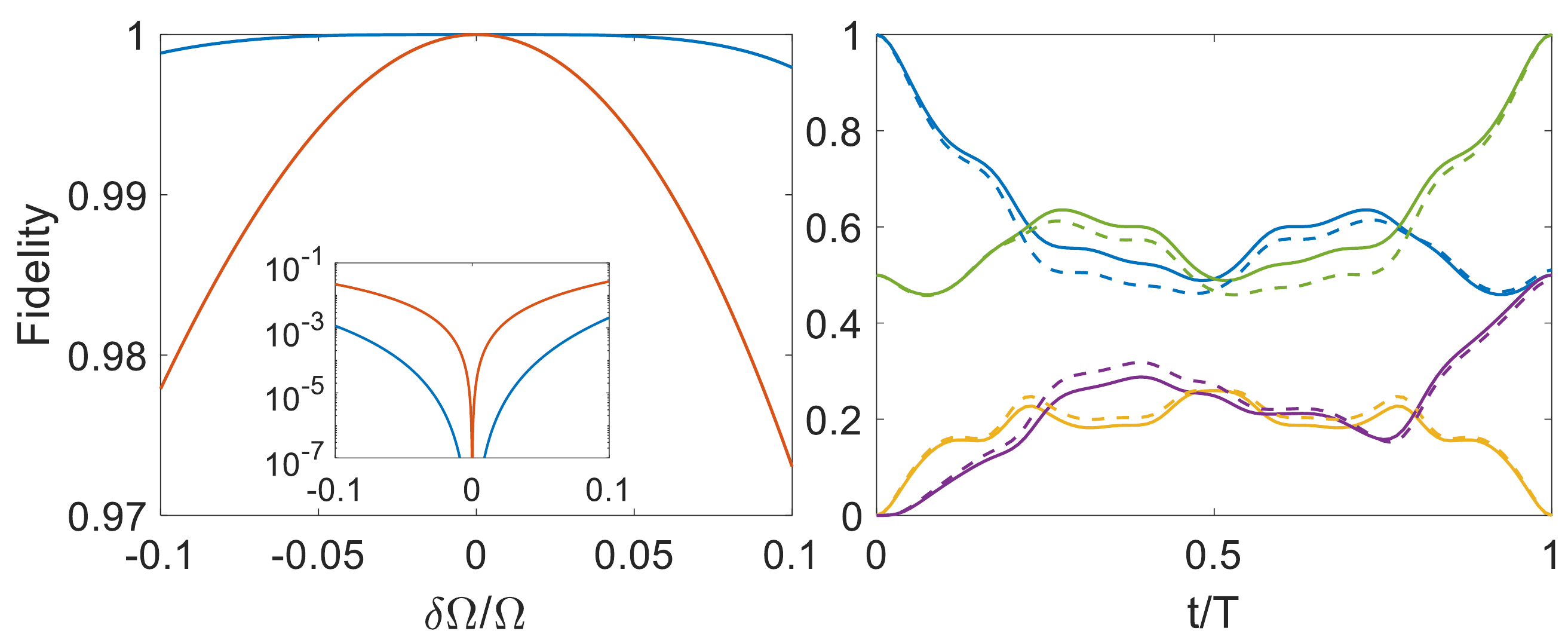} \caption{Minimal gate scheme. Analogous to Fig. 1 of the main text. Left: Fidelity of the gate in presence of deviations in the Rabi frequency, $\delta\Omega$. The minimal gate shows a high-fidelity and wide repsponse, similar to the gate shown in the main text. Right: Population dynamics of the gate for the initial state $\ket{00}$ in the motional ground state for the ideal (solid) and errorneous, $\delta\Omega/\Omega=5\%$ (dashed) cases. Both exhibit a high fidelity (green) operation at $t=T$ as the $\ket{00}$ (blue) and $\ket{11}$ (purple) populations are equal and the $\ket{01}$ and $\ket{10}$ (orange) vanish. This gate is not robust to timing errors, therefore the dynamics around the gate time change quadratically.}
\label{sfigMain}
\end{figure}

\begin{figure}
\centering{}\includegraphics[width=0.5\columnwidth]{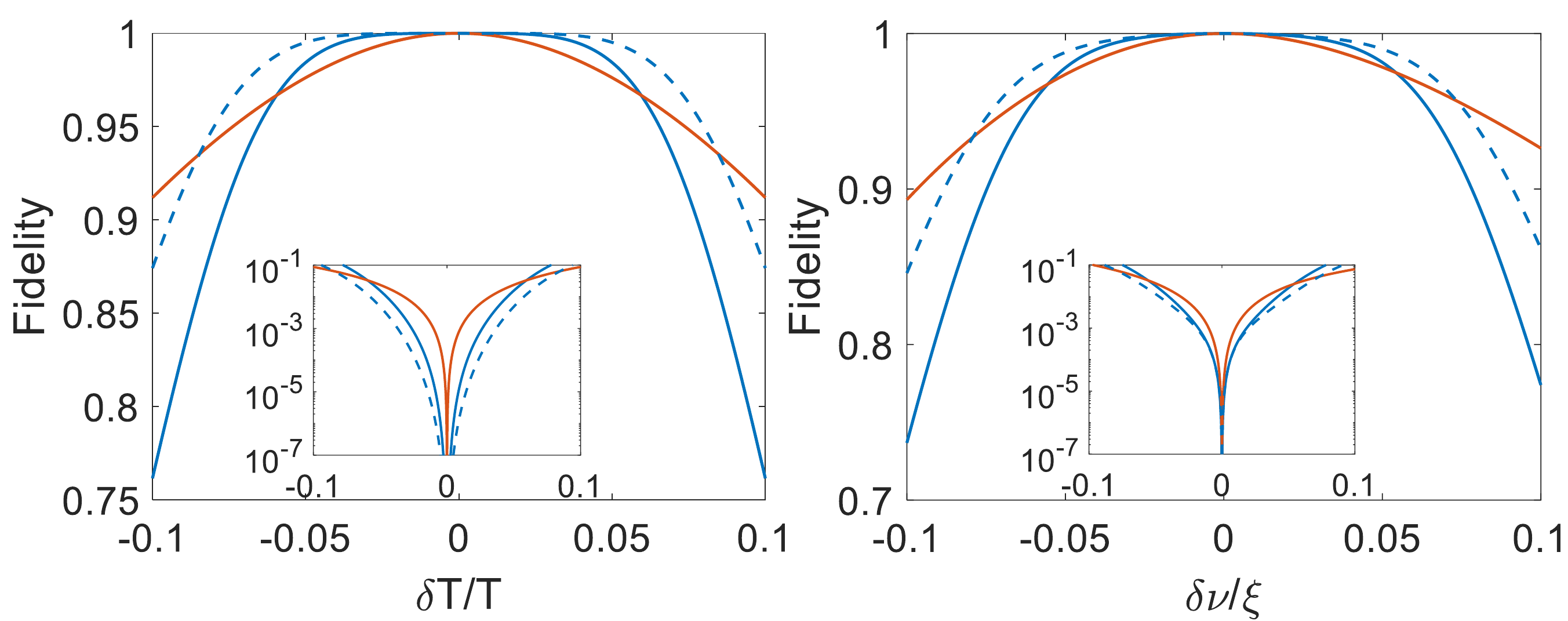} \caption{Minimal gate scheme. Analogous to Fig. 4 of the main text comparing the minimal gate (blue) with the MS gate (red) and the robust gate in the main text (dashed blue). Left: Gate fidelity in presence of timing errors. Right: Gate fidelity in presence of motional  frequency errors. For both cases the minimal gate exhibit an improved response upon the MS gate in a finite range of error, this is due to the fact that its sideband is driven with $a_3$. The robust gate from the main text exhibits a superior reponse.}
\label{sfigFid2}
\end{figure}

This gate's performance is presented in Figs. \ref{sfigMain} and \ref{sfigFid2}. Specifically, Fig. \ref{sfigMain}, analogous to Fig. 1 of the main text, shows the fidelity (left) of the minimal gate (blue) compared to the MS gate (red). Clearly the gate exhibits increased robustness to deviation in $\Omega$. The population dynamics are shown (right) for the ideal case (solid) and a erroneous, $\delta\Omega/\Omega=5\%$ case. We observe a high-fidelity entangling gate for both cases at $t=T$. 

Figure \ref{sfigFid2}, analogous to Fig. 4 of the main text, shows the fidelity of the gate for gate timing errors (left) or motional frequency errors (right). In both we compare the minimal gate (blue) with the MS gate (red) and the robust gate in the main text (dashed). The minimal gate shows an improved response, compared to the MS. This can be understood by the fact that the first motional sideband is driven with its third harmonic, i.e. with $a_3$ \cite{roos2008ion}. Nevertheless, since this gate is not actually designed for robustness it is still less resilient than the scheme we present in the main text and in the next subsection.

\section{Additional robustness constraints}
Next, we modify the minimal gate above such that it would have additional robustness properties, using constraints that were developed previously. Specifically, those include robustness to timing errors, motional mode heating, trapping frequency errors, and off-resonance coupling to unwanted transitions. These are provided by the constraints \cite{haddadfarshi2016high,shapira2018robust,webb2018resilient,bentley2020numeric},
\begin{align}
    w_1\left(t=0\right) =& w_1\left(t=T\right)=0, \\
    w_2\left(t=0\right) =& w_2\left(t=T\right)=0, \\
    \partial_t \alpha |_{t=T}=&0, \\
    \partial_t\left(\Phi_2+\Phi_4\right)|_{t=T}=&0,\\
    \partial_t\Phi_3|_{t=T}=&0,\\
    \left\{\alpha\right\}=0.\label{eqMoreCons}
\end{align}

We note that satisfying the first two constraints makes all other constraints redundant, except the last. The first two constraints are satisfied by setting $\sum_n a_{2n+1}=0$ and $\sum_n n  s_{2n}=0$. The last constraint is valid in leading order in $r$ and therefore we satisfy it to that order, yielding $\sum_n a_{2n+1}/\left(2n+1\right)=0$. All of these constraints are linear in the $a_n$'s and $s_n$'s and are thus easy to implement, yielding the drive in Eq. (15) of the main text.

\section{Analytic approximate solution of phase constraints}
The constraints (C5) and (C6) are first solved approximately by expanding the constraints to sixth order in the driving field's amplitudes, $a$ and $s$, and solving the expanded constraints analytically. Then, we numerically optimize the results using a local gradient descent, over the full expressions of the constraints.

The sixth order expansion of (C5) and (C6) yields,
\begin{align}
\frac{4\pi a^{2}}{135135\xi^{2}}\left(71582s^{4}+106431s^{2}-225225\right)&=-\frac{\pi}{2}\\
\frac{8\pi a^{2}}{45045\xi^{2}}\left(71582s^{4}+70954s^{2}-75075\right)&=0,\label{eqExpConstraints}
\end{align}
respectively. We first solve the expanded version of Eq. (C6), as it is a simply the quadratic root in $s^{2}$. We then plug the solution in the expanded (C5), yielding our approximation $a\approx-0.364172\xi$ and $s\approx-0.801322$. This approximation is a sufficient starting point for the local numerical solution, which quickly converges to the full solution in the main text.

\section{Power requirements}

\begin{figure}
\centering{}\includegraphics[width=0.5\columnwidth]{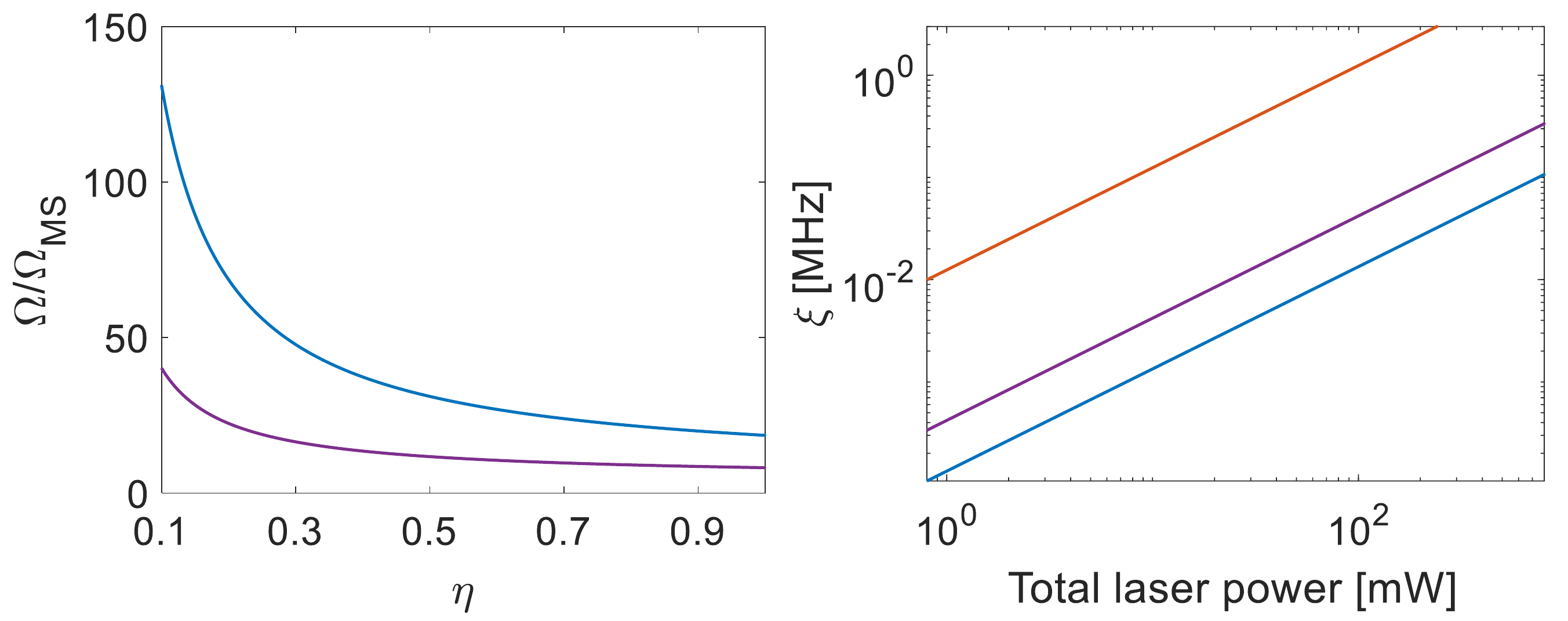} \caption{Driving amplitude requirements. Left: Drive amplitude of the robust (blue) and minimal (purple) gate, normalized by the MS drive amplitude, as a function of the Lamb-Dicke parameter, $\eta$. Right: Gate rate, $\xi$, as a function of total laser intensity for the specific realization considered in the text, for the robust (blue), minimal (purple) and MS (red) gates.}
\label{sfigPower}
\end{figure}

We consider the power requirements of our robust gate scheme as well as the minimal robust gate scheme, compared to the MS gate. The ratio $\Omega/\Omega_\text{MS}$ is independent of details of implementation and is shown in Fig. \ref{sfigPower} (left) for the robust gate in the main text (blue) and the minimal robust gate (purple), both showing a $1/\eta$-like scaling which is due to the second sideband drive. 

We consider a specific, yet conventional, realization. Namely two $^{40}\text{Ca}^+$ ions, driven by a $400\text{ nm}$ laser in a counter-propagating Raman transition, with a beam waist, $\omega_0=35\um.$ The ions are coupled to a center-of-mass normal-mode of motion with frequency $3\text{ MHz}$. This results with $\eta\approx0.144$, such that $\Omega_\text{robust}/\Omega_\text{MS}\approx68.6$ and $\Omega_\text{minimal}/\Omega_\text{MS}\approx22.5$. Furthermore, we plot the the gate rate $\xi=2\pi/T$ vs. the total laser intensity illuminating the ions, shown in Fig. \ref{sfigPower} (right) for the robust gate (blue), the minimal gate purple) and MS (red). For a multi-ion chain the required intensity can be steered to exclusively illuminate the two target ions using, for example, a microelectromechanical system \cite{wang2020high} or an acousto-optic deflector \cite{manovitz2022trapped}.

\section{Fidelity calculation}
We define the fidelity of the gate as,
\begin{align}
F&=\left|\broket{00;n=0}{U_{\text{ideal}}^{\dag}U_{I}}{00;n=0}\right|^{2},
\end{align}
with $U_{\text{ideal}}=\exp\left(i\pi J_{x}^{2}/2\right)$, $\ket{00}$ the two-qubit ground state and $\ket{n=0}$ the motional ground state. That is we compare the overlap of the ideal operation of our gate with its actual operation, assuming the initial condition is the qubit and motional ground state. 

Deviations such as $\delta\Omega$, $\delta T$ and $\delta\nu$ can be readily inserted into the expression of $U_{I}$ in Eq. (7) of the main text. 

We evaluate F directly,
\begin{align}
F	=&\left|\broket{00;0}{e^{i\frac{\pi}{2}J_{x}^{2}}S\left(J_{x}r\right)D\left(\alpha\right)e^{-i\left(J_{x}^{2}\left(\Phi_{2}+\Phi_{4}\right)+J_{x}\Phi_{3}\right)}}{00;0}\right|^{2}\\
=&\left|\broket{00;0}{S\left(J_{x}r\right)D\left(\alpha\right)e^{-i\left(J_{x}^{2}\left(\Phi_{2}+\Phi_{4}-\frac{\pi}{2}\right)+J_{x}\Phi_{3}\right)}}{00;0}\right|^{2}.
\end{align}

Using, $\ket{00}=\frac{1}{2}\left(\ket{++}+\ket{+-}+\ket{-+}+\ket{--}\right)$, where $+$ and $-$ indicate the $\sigma_x$ eigenvalues of the two spins, and since the only spin operators occurring in the expression for $F$ are $J_{x}$, we have,
\begin{align}
F	=&\left|\frac{1}{2}+\frac{1}{4}e^{-i\left(\left(\Phi_{2}+\Phi_{4}-\varphi\right)+\Phi_{3}\right)}\broket 0{S\left(r\right)D\left(\alpha_{+}\right)}0+\frac{1}{4}e^{-i\left(\left(\Phi_{2}+\Phi_{4}-\frac{\pi}{2}\right)-\Phi_{3}\right)}\broket 0{S\left(-r\right)D\left(\alpha_{-}\right)}0\right|^{2}\\
=&\left|\frac{1}{2}+\frac{1}{4}e^{-i\left(\left(\Phi_{2}+\Phi_{4}-\varphi\right)+\Phi_{3}\right)}\broket 0{D\left(\gamma_{+}\right)S\left(r\right)}0+\frac{1}{4}e^{-i\left(\left(\Phi_{2}+\Phi_{4}-\varphi\right)-\Phi_3\right)}\broket 0{D\left(\gamma_{-}\right)S\left(-r\right)}0\right|^{2},
\end{align}
with,
\begin{align}
\gamma_{+}=&\alpha_{+}\cosh\left(r\right)-\alpha_{+}^{\ast}\sinh\left(r\right)\\
\gamma_{-}=&\alpha_{-}\cosh\left(r\right)+\alpha_{-}^{\ast}\sinh\left(r\right),
\end{align}
and $\alpha_{+}$ ($\alpha_{-}$) the phase-space trajectory associated with the state $\ket{++}$ ($\ket{--}$).

Using \cite{gerry2004introductory},
\begin{equation}
\brkt{0}{\alpha,\xi}=\broket {0}{D\left(\alpha\right)S\left(\xi\right)}{0}=\frac{\exp\left[-\frac{1}{2}\left|\alpha\right|^{2}-\frac{1}{2}\alpha^{\ast2}e^{i\theta}\tanh r\right]}{\cosh\left(r\right)},
\end{equation}
we get,
\begin{align}
F	=&\left|\frac{1}{2}+\frac{1}{4}\frac{e^{-i\left(\Phi_{2}+\Phi_{4}-\frac{\pi}{2}\right)}}{\cosh\left(r\right)}\left(\exp\left[-i\Phi_{3}-\frac{1}{2}\left|\gamma_{+}\right|^{2}-\frac{1}{2}\gamma_{+}^{\ast2}\tanh r\right]+\exp\left[i\Phi_{3}-\frac{1}{2}\left|\gamma_{-}\right|^{2}+\frac{1}{2}\gamma_{-}^{\ast2}\tanh r\right]\right)\right|^{2},
\end{align}
where the implicit expressions for $\Phi_{2,3,4}$ are in Eq. (9)-(11) of the main text and $\gamma_\pm$ are defined above. We note that for an ideal gate at the gate time $\Phi_{3}=\gamma_{+}=\gamma_{-}=0$ and we get, $F=1$ as expected.

\bibliography{main}